\documentclass[11pt]{article}

\usepackage{varioref}
\usepackage{amsmath,amsfonts,amsthm,mathrsfs}

\usepackage[normalem]{ulem}
\usepackage{multirow,color,graphics}
\usepackage{amsmath}
\usepackage{amsfonts}
\usepackage{amssymb}
\usepackage{jheppub}
\usepackage{enumerate}
\usepackage{fancyvrb}
\usepackage{verbatim}
\usepackage{wrapfig}
\usepackage{appendix}
\usepackage{amstext}
\usepackage{amssymb}
\usepackage{graphicx}
\usepackage{color}
\usepackage{varioref}
\usepackage{multirow,graphics}
\usepackage{epstopdf}
\newcommand{\nn}{\nonumber}

\numberwithin{equation}{section}

\newcommand{\hl}{{\hat \lambda}}

\def\[{\left[}
\def\]{\right]}
\def\({\left(}
\def\){\right)}

    \newcommand{\beq}{\begin{equation}}
    \newcommand{\eeq}{\end{equation}}
    \newcommand\beqa{\begin{eqnarray}}
    \newcommand\eeqa{\end{eqnarray}}
\newcommand\bea{\begin{array}}
\newcommand\eea{\end{array}}

\newcommand{\la}[1]{\label{#1}}
\newcommand{\eq}[1]{(\ref{#1})}

\makeatletter
     \@ifundefined{usebibtex}{} {}
\makeatother

\newcommand{\cO}{{\cal O}}

\title{On the Exact Interpolating Function in ABJ Theory}

\author[]{~Andrea Cavagli\`{a},${}^{1}$}
\author[]{~Nikolay Gromov,${}^{2,3}$}
\author[]{~Fedor Levkovich-Maslyuk${}^{2}$}

\affiliation[]{${}^1$ Dipartimento di Fisica and INFN, Universit\`{a} di Torino, Via P. Giuria 1, 10125 Torino,
Italy }
\affiliation[]{${}^2$Mathematics Department, King's College London,
The Strand, London WC2R 2LS, UK}
\affiliation[]{${}^3$St.Petersburg INP, Gatchina, 188 300, St.Petersburg,
  Russia}

\emailAdd{cavaglia$\bullet$to.infn.it}
\emailAdd{nikgromov$\bullet$gmail.com}
\emailAdd{fedor.levkovich$\bullet$gmail.com}

\abstract{
Based on the recent indications of integrability in the planar ABJ model,
we conjecture an exact expression for the interpolating function $h(\lambda_1,\lambda_2)$ in this theory.
Our conjecture is based on the observation that the integrability structure of the ABJM theory given by its Quantum Spectral Curve
is very rigid and does not allow for a simple consistent modification.
Under this assumption, we revised the previous comparison of localization
results and exact all loop integrability calculations done for the ABJM theory
by one of the authors and Grigory Sizov, fixing $h(\lambda_1,\lambda_2)$. We checked our conjecture against various
weak coupling expansions, at strong coupling and also demonstrated its invariance under the Seiberg-like duality.
This match also gives further support to the integrability of the model.
If our conjecture is correct, it extends all the available integrability results in the ABJM model to the ABJ model.
}

\begin{document}

\maketitle

\section{Introduction}
Integrability of AdS/CFT has a long history and over the past few years gave numerous exciting results.
First discovered in ${\cal N}=4$ SYM in 4D \cite{Minahan:2002ve}, the integrability methods have been exported to the Aharony-Bergman-Jafferis-Maldacena (ABJM) theory in 3D  (proposed in \cite{Aharony:2008ug} following \cite{Gustavsson:2007vu,Bagger:2007jr})
and to some 2D theories as well in the planar limit, see the reviews \cite{Beisert:2010jr,Sfondrini:2014via}. Currently the planar spectrum in many of these examples seems to be under complete control of integrability but also numerous results are available for Wilson loops, amplitudes and even 3-point correlators.

The ABJ model \cite{Aharony:2008gk} is a simple generalization of the 3D ${\cal N}=6$ ABJM Chern-Simons-matter theory
and its gauge group is $U(N_1)_k \times U(N_2)_{-k}$ where $k$ indicates the Chern-Simons level.
In the planar limit this theory has two `t Hooft couplings $\lambda_1=N_1/k$ and $\lambda_2=N_2/k$,
and for the particular case  $\lambda_1=\lambda_2$ we get the ABJM model.
The ABJ theory has a well established AdS dual, but whether or not it is integrable has been unclear for a long time.
One of the reasons for this is that the string dual for the ABJ model is indistinguishable from that of the ABJM model
to all orders in perturbation theory. At the same time, calculations at weak coupling are highly
complicated but also revealed rather slim differences with the ABJM case. Namely, in every situation studied so far it was possible to absorb all the dependence on two separate couplings into a redefinition of the `t Hooft coupling of the ABJM model \cite{Bak:2008vd,Minahan:2009te,Minahan:2009aq,Minahan:2009wg,Leoni:2010tb}.

One of the possible approaches to the solution of this theory, which we adopt here, would be to assume
integrability and try to draw some predictions from that. One should be warned at this point that
the $\theta$-term present in the worldsheet string theory \cite{Aharony:2009fc} could break integrability, similarly to what may happen in some non-supersymmetric sigma models.
Nevertheless we can
play this game,
and in particular we will be able to make a prediction for the single magnon dispersion relation which
passes all tests presently available.

Discussing integrability in ABJ(M) theory it is instructive to use the historical path for a moment. First integrability was
developed in the asymptotic limit for very long single trace operators/strings, the construction being based on the S-matrix which
is (up to a scalar factor) fixed completely by the symmetry \cite{Minahan:2008hf,Gromov:2008qe,Ahn:2008aa}. The global symmetry of both ABJ and ABJM theories is the same, which
suggests
that the integrable
structure, if it exists, is likely to only differ by a redefinition of the coupling constant, as already pointed out in \cite{Bak:2008vd,Minahan:2009te}. For short operators the S-matrix approach becomes unreliable and one should use the Y-system/TBA \cite{Bombardelli:2009xz,Gromov:2009at} or its reformulation as the Quantum Spectral Curve (QSC)
which is a simple set of Riemann-Hilbert equations. The QSC was first formulated in ${\cal N}=4$ SYM \cite{Gromov:2013pga,Gromov:2014caa} and later extended to the ABJM model in \cite{Cavaglia:2014exa}. In particular in \cite{Cavaglia:2014exa} it was noticed that
the QSC for the ABJM theory algebraically has exactly the same form as in ${\cal N}=4$ SYM, suggesting its structure is rather rigid.
Indeed, we found it complicated
to modify the QSC construction to incorporate
an extra parameter.
The conclusion is that it would be hardly possible
to include into the construction two different `t Hooft couplings in any way except replacing $\lambda\to \lambda^{\text{eff}}(\lambda_1,\lambda_2)$.

A key observable in the theory is the magnon dispersion relation representing a simplest perturbation of a long BPS operator.
Symmetry constrains it to be of the form
\beq
	\gamma(p)=\sqrt{\frac{1}{4}+4h^2(\lambda_1,\lambda_2)\sin^2\frac{p}{2}}-\frac{1}{2}
\eeq
so that it is given in terms of an interpolating function $h(\lambda_1,\lambda_2)$. At the same time this interpolating function
determines the positions of the branch points in the QSC construction, where they are situated at $\pm 2 h$.
Thus $h$ is a physical observable of the theory which plays a central role in the integrability construction.
In the next section we describe our conjecture for this quantity and then go through some tests of our proposal.

\section{Conjecture for the interpolating function}
The conjecture for the expression of $h(\lambda_1,\lambda_2)$ which we put forward in this note is
\beq
\label{hab0}
	h(\lambda_1,\lambda_2)=\frac{1}{4\pi}\log\left(\frac{a b+1}{a+b}\right)\;.
\eeq
where $a$ and $b$ (which we assume $|a|\geq 1$ and $|b|\geq 1$) parameterize $\lambda_1$ and $\lambda_2$ in the following way
\beq\la{intl}
	\lambda_1=-\frac{1}{4\pi^2}\oint_{a}^{1/a}\omega(Z)\frac{dZ}{Z}\;\;,\;\;
	\lambda_2=+\frac{1}{4\pi^2}\oint_{-b}^{-1/b}\omega(Z)\frac{dZ}{Z}\;,
\eeq
and where
\beq
	\omega(Z)=\log\(\sqrt{(Z+b)(Z+1/b)}-\sqrt{(Z-a)(Z-1/a)}\)\;.
\eeq
The map $(a,b)\to(\lambda_1,\lambda_2)$ is not single-valued as we shall discuss later in relation to a Seiberg-like duality.
However, for small enough $\lambda_1,\lambda_2$ it gives unique mapping between the branch points and the couplings.

For some observables, accessible by the localization, the points $a,1/a,b,1/b$
have the meaning of the end of the cuts on which the roots of the matrix model condense
as we also discuss below.

The main inspiration for our conjecture comes from the calculation of \cite{Gromov:2014eha} where $h$ was fixed in the ABJM case by comparing a localization prediction with an integrability-based result. For the case $\lambda_1=\lambda_2=\lambda$
the integrals can be solved explicitly \cite{Marino:2009jd} and our
 result for $h(\lambda_1,\lambda_2)$
reduces to the one of \cite{Gromov:2014eha}
\beq
	\lambda = \frac{\sinh(2\pi h)}{2\pi}{}_3F_2\left(\frac{1}{2},\frac{1}{2},\frac{1}{2};1,\frac{3}{2};
-\sinh^2(2\pi h(\lambda,\lambda))\right)\;.
\eeq
This will be discussed in more detail in section 3.

We now describe various tests of our conjecture and make a comparison with the known results
for $h(\lambda_1,\lambda_2)$ in several limits.

\subsection{Reality and analyticity}
A first test of our formula is that the result for $h(\lambda_1,\lambda_2)$ is real in the physical range of the parameters, namely for $\lambda_1,\, \lambda_2 \geq 0$ with $|\lambda_1-\lambda_2| < 1 $ \cite{Aharony:2008gk}.
In fact this is already rather nontrivial, as the points $a$ and $b$ are complex numbers
with no obvious conjugation symmetry. In order to prove reality it is convenient to parameterize $a,b$ in terms of the new variables $B$ and $\kappa$ used in \cite{Drukker:2010nc} and defined by
\beq
\label{abkB}
	4e^{2\pi i (B-1/2)}=a+\frac{1}{a}+b+\frac{1}{b}\ ,\ \ \ 2\kappa e^{\pi i B}=a+\frac{1}{a}-b-\frac{1}{b}\ .
\eeq
As shown in \cite{Drukker:2010nc}, both $B$ and $\kappa$ are real, and in fact
\beq
\label{Blam}
	B=\lambda_1-\lambda_2+1/2\ .
\eeq
Moreover, we have
\beq
\label{kaineq}
	\kappa \geq 4|\cos \pi B|\ .
\eeq
In these variables the expression \eq{hab0} for $h(\lambda_1, \lambda_2)$ takes the form
\beq
\label{hkaB}
	h(\lambda_1, \lambda_2)=\frac{1}{4\pi}\log\( u + \sqrt{u^2 -1} \)\ ,
\eeq
where
we defined
\beq\label{eq:u}
	u= \frac{\kappa^2}{8}-\cos2\pi B\ .
\eeq
From \eq{kaineq} it follows that $u\geq 1$, so the expression inside the logarithm in \eq{hkaB} is real and greater than 1, and therefore $h(\lambda_1,\lambda_2)$ is indeed real and positive.

Another interesting hint at the correctness of our result is the correspondence between singular points in the matrix model construction and the structural properties of the QSC. In particular, the matrix model description becomes singular at the so-called conifold locus \cite{Drukker:2010nc}, where either $a=\pm1$ or $b=\pm 1$, and consequently one of the branch cuts collapses. In terms of the interpolating function (\ref{hab0}), such points correspond either to $h(\lambda_1, \lambda_2) \rightarrow 0$, or to the complex points $h(\lambda_1, \lambda_2) \rightarrow i/4 $ (modulo the $i /2$ periodicity due to the $\log$ in (\ref{hab0})). We would like to point out that the points $h = \pm i/4$ are indeed special ponts in the QSC formulation, where the branch points in the rapidity plane  touch each other, leading to the formation of a singularity as is the case also for $\mathcal{N}=4$ SYM.
The correspondence with the matrix model description yields another indirect confirmation of   our proposal.

\subsection{Weak coupling limit}
At weak coupling one has $a\sim b\sim 1+{\cal O}(\sqrt{\lambda_a})$ so the contour integral can be expanded first
for $a,b\sim 1$ and then computed by residues (similar expansions in this setup were studied in \cite{Aganagic:2002wv,Marino:2009jd,Drukker:2010nc}, and are known as the orbifold limit of the matrix model). This leads to the following expression for $h(\lambda_1, \lambda_2)$:
\beqa\label{eq:hweak}
h^2(\lambda_1, \lambda_2) &=& \lambda_1 \lambda_2 -\frac{\pi ^2}{6}  \lambda_1 \lambda_2 \left( \lambda_1 + \lambda_2 \right)^2 \\
&& \nn +\frac{\pi ^4}{360}  \lambda_1 \lambda_2 \left( \lambda_1 + \lambda_2 \right)^2 \, \left( 3 \lambda_1^2 + 3 \lambda_2^2 + 79 \, \lambda_1 \lambda_2 \right) \\
&& \nn -\frac{\pi^6}{15120} \, \lambda_1 \lambda_2 \left( \lambda_1 + \lambda_2 \right)^2 \,\left( 3 \lambda_1^4 + 533 \lambda_1^3 \lambda_2 +
 5336 \lambda_1^2 \lambda_2^2 + 533 \lambda_1 \lambda_2^3 + 3 \lambda_2^4 \right)
\\ \nn &&
  + \cO\left( |\lambda_k|^{10}\right)\ .
	\nn
\eeqa
This expansion can be compared against the direct 4-loop perturbative calculation
of \cite{Minahan:2009aq,Minahan:2009wg} (later also confirmed in \cite{Leoni:2010tb}), which gives
\beq
	h_{4-loop}^2(\lambda_1,\lambda_2)=\lambda_1\lambda_2+h_4\(\lambda_1\lambda_2\)^2+h_{4,\sigma}\lambda_1\lambda_2(\lambda_1-\lambda_2)^2
\eeq
where
\beq
	h_4=-\frac{2\pi^2}{3},\ \ h_{4,\sigma}=-\frac{\pi ^2}{6}\;,
\eeq
perfectly matching our result! Thus our result reproduces all three known coefficients in the perturbative expansion (one coefficient at the leading order and two at the next order).

A part of the 6-loop dilatation operator in ABJ theory was computed in \cite{Bak:2009tq}.
It would be interesting to complete this calculation and obtain 6-loop anomalous dimensions to make a more detailed comparison at weak coupling. Weak coupling results based on the QSC formulation were obtained in \cite{Anselmetti:2015mda}.

\subsection{Partial weak coupling limit}
Another interesting limit where we can make a comparison with known data is the limit where only one of the `t Hooft couplings
goes to zero. In this limit (which is in fact an expansion near the conifold locus with $a \sim 1$ or $b\sim 1$) again the integrals \eq{intl} can be solved analytically, resulting in the following
expression
\beqa\label{eq:hextr}
h^2(\lambda_1, \lambda_2) &=&\frac{\lambda _1 }{\pi
   }\sin \left(\pi  \lambda _2\right)+\frac{ \lambda _1^2}{3}\sin ^2\left(\frac{\pi
   \lambda _2}{2}\right) \left(1-5
   \cos \left(\pi  \lambda _2\right)\right)\\
   &+&\frac{\pi\lambda _1^3}{720}
    \left(-639   \sin \left( \pi  \lambda
   _2\right)+600  \sin \left(2 \pi  \lambda
   _2\right)-227  \sin \left(3 \pi  \lambda
   _2\right)\right)+{\cal O}\left(\lambda _1^4\right)\nn\;.
\eeqa
This limit was considered in  \cite{Minahan:2010nn} where a conjecture was put forward for the dilatation operator at leading order in $\lambda_1$ but to all loops in $\lambda_2$. Very recently
that conjecture was confirmed
in \cite{Bianchi:2016rub} and put on firmer grounds. Remarkably, the prediction for the leading order in $\lambda_1$ given in these papers
precisely matches the first term in the expansion above! This provides another nontrivial test of our conjecture, to all orders in $\lambda_2$.

\subsection{Strong coupling}
We consider now the strong coupling regime in which $\lambda_1,\lambda_2 \rightarrow \infty$, while $\lambda_1-\lambda_2$ stays finite. Notice that this is the generic physical strong coupling region since unitarity was argued in \cite{Aharony:2008gk} to require $|\lambda_1 - \lambda_2| \leq 1$. This limit was studied in detail in  \cite{Drukker:2010nc} and it is convenient to switch from $a,b$ to $\kappa$ and $B$ defined by \eq{abkB}. Due to \eq{Blam}, $B$ is finite in this regime.
The strong coupling expansion of $\kappa$ is then arranged as the convergent expansion \cite{Drukker:2010nc}
\beqa\label{eq:strc}
\kappa &\sim& e^{\pi \sqrt{2 {\hat \lambda } } } \, \left( 1 + \sum_{l \geq 1} c_l\left( x, y \right) \,
e^{-2 l \pi \sqrt{2 \hl } } \right)  ,
\eeqa
with $x=(\pi\, \sqrt{2 \hl } )^{-1} $, $y = -e^{2 \pi i B}  $,
where we see the appearance of the redefined t'Hooft parameter
\beqa
{\hat \lambda} \equiv \frac{\lambda_1 + \lambda_2 }{2}-\frac{1}{2}\left(B - \frac{1}{2} \right)^2 - \frac{1}{24} = \frac{\lambda_1 + \lambda_2 }{2}-\frac{1}{2}\left(\lambda_1-\lambda_2\right)^2 - \frac{1}{24} ,
\eeqa
which from string theory arguments is expected to be the natural variable at strong coupling \cite{Drukker:2010nc,Aharony:2009fc}. The coefficients $c_l(x, y)$ appearing in (\ref{eq:strc}) can be written as polynomials in $x$ and $\cos(2 \pi m B)$, with $m=0, \dots, l$.
Using the first two coefficients computed in \cite{Drukker:2010nc} it is simple to determine the strong coupling behaviour of our proposal for $h(\lambda_1, \lambda_2)$,
\beqa
\label{hstrongnp}
&&h(\lambda_1, \lambda_2) =
 \sqrt{ \frac{{\hat \lambda}}{2}} - \frac{\log(2)}{2 \pi}  - \frac{ e^{-2 \pi \sqrt{2 {\hat \lambda} }} \, \cos( 2 \pi B)}{\pi} \, \left(1 + \frac{1}{\pi \, 2\, \sqrt{ 2 \hl} } \right) \\
\nn && \;\;- \frac{ e^{-4 \pi \sqrt{2 {\hat \lambda} }} }{2\pi} \, \left(6 + \cos(4 \pi B)  + \frac{16 + 9 \cos( 4 \pi B) }{4 \pi \sqrt{2 \hl} } + \frac{\cos(2 \pi B)^2 }{\pi^2 \, \hl } + \frac{\cos(2 \pi B)^2 }{4 \sqrt{2} \, \pi^3  \, \hl^{\frac{3}{2} } }\right)
\\ \nn &&
 + \mathcal{O}\left(e^{-6 \pi \sqrt{2 {\hat \lambda} }}\right) .\nn
\eeqa
As we see the result written in terms of the shifted $\hat\lambda$
is indistinguishable from that in the ABJM model to all orders in perturbation theory in the natural world-sheet coupling $\hat \lambda$. It would be interesting to try
to extricate the nonperturbative terms in \eq{hstrongnp} from a first principles worldsheet calculation.

\subsection{Symmetry}

The function $h(\lambda_1,\lambda_2)$ is expected to be invariant under the transformation
\beq
\label{seitrans}
	(\lambda_1,\lambda_2)\to (2\lambda_2-\lambda_1+1,\lambda_1)
\eeq
which corresponds to a Seiberg-like duality linking ABJ theories with two different gauge groups,
$U(N_1)_k\times U(N_2)_{-k}$ and $U(2N_2-N_1+k)_k\times U(N_1)_{-k}$ \cite{Aharony:2008gk}. Using the matrix model arising from localization, this duality was proved for some observables at finite $N_1,N_2$ in \cite{Kapustin:2010mh, Awata:2012jb, Honda:2013pea}. Let us briefly discuss how this symmetry
 manifests itself in the planar limit and show that our result is invariant under it\footnote{Its effect at strong coupling was also discussed in \cite{Drukker:2010nc}.}.

\begin{figure}[h]
\begin{center}
\includegraphics[scale=0.6]{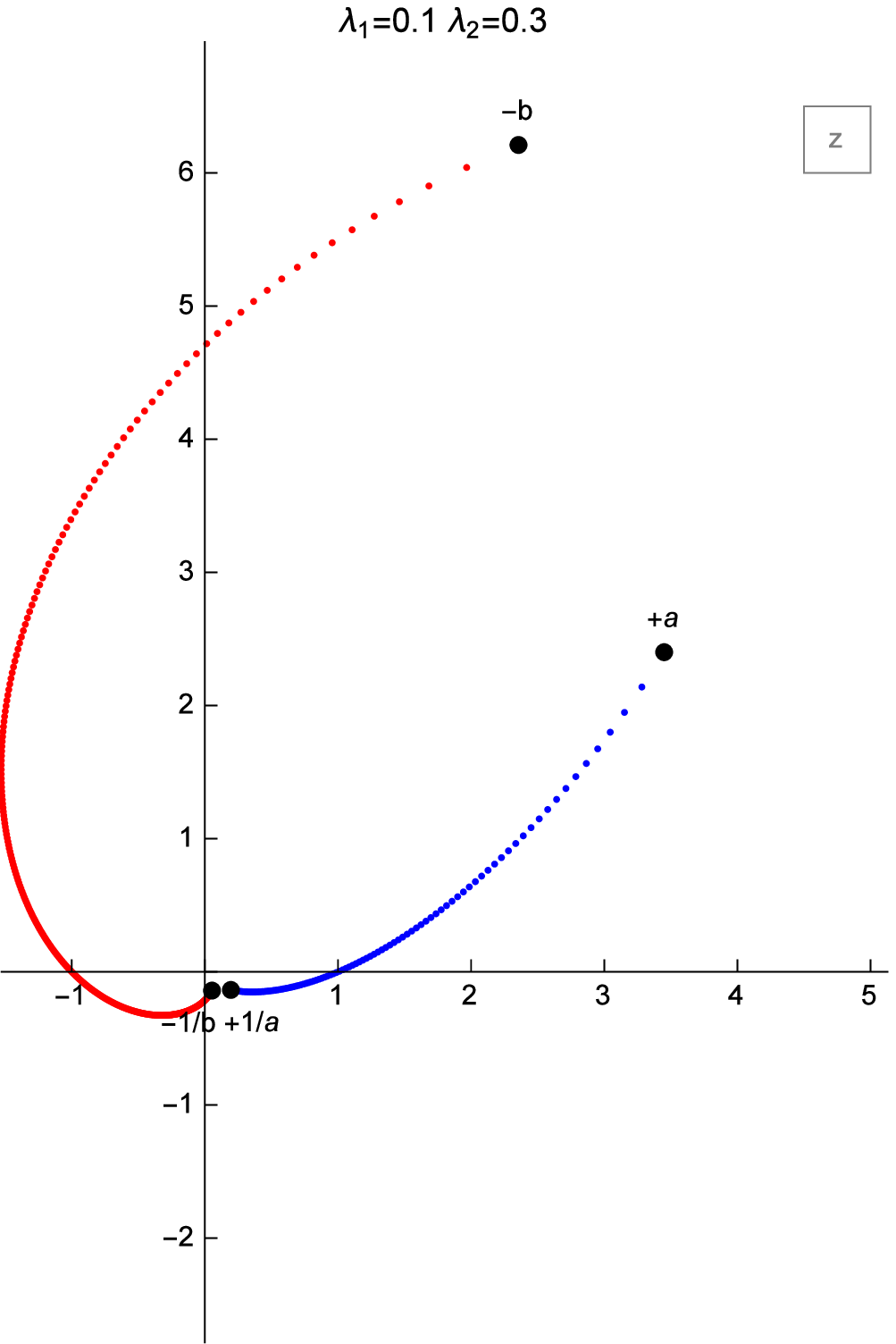}
\includegraphics[scale=0.6]{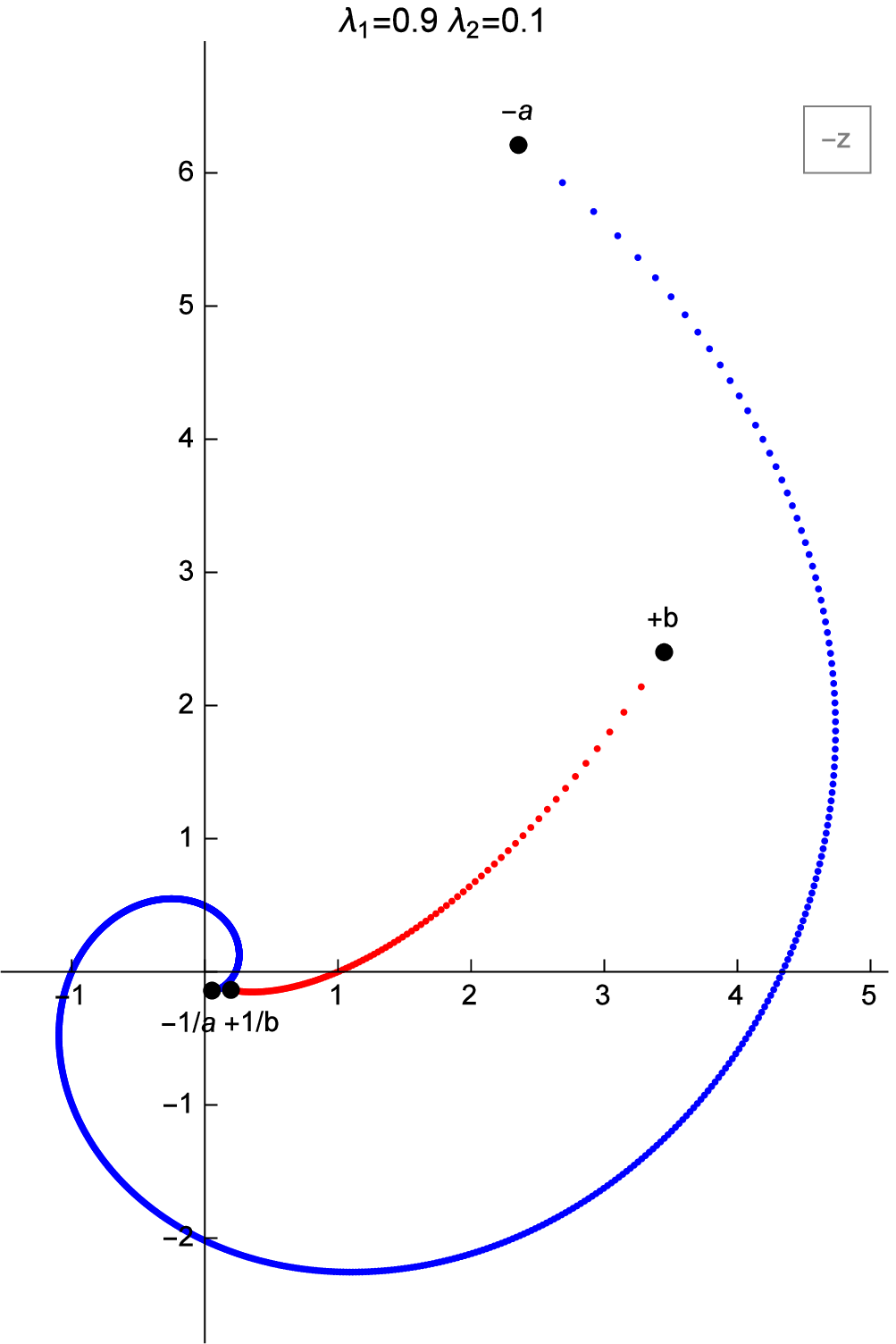}
\caption{Numerical solution of the ABJ matrix model for two sets of dual couplings.
We see that the branch points are exactly the same (up to a sign), while for the second configuration
one of the cuts encircles the origin
and another cut. This results in an extra contribution $+1$ from the pole at the origin and
$2\lambda_2$ from the second (red) cut which results in $2\lambda_2-\lambda_1+1$ for the total integral around the new contour.
As the branch points simply flip their signs and interchange $a\leftrightarrow b$
the interpolating function remains invariant.
\label{fig:symm}
}
\end{center}
\end{figure}

In fact the transformation \eq{seitrans} can be easily understood by a simple rearranging of the integration contours.
To understand exactly how that works we solved numerically the underlying discrete matrix model saddle point equation
for some large number of roots ($\sim 1000$) for two sets of $(\lambda_1,\lambda_2)$ related by the symmetry.
The distribution of the roots is given in Fig. \ref{fig:symm}.

The way the symmetry works is that the parameters get mapped as $(a,b)\to(b,a)$.
Whereas one of the integration contours remains unchanged (up to a sign flip), the second one winds around the origin
and another cut. The pole at the origin gives $+1$ as an extra contribution and the second encircled cut
gives twice $\lambda_2$, resulting in the transformation $(2\lambda_2-\lambda_1+1,\lambda_1)$.
We notice that the transformation $(a,b)\to(b,a)$ keeps $h(\lambda_1, \lambda_2)$ unchanged ensuring the symmetry of that quantity.

Note that one can start from ABJM theory where $\lambda_1=\lambda_2$ and generate predictions
for $h(\lambda_1,\lambda_2)$ with non-equal couplings. It is quite notable that in our proposal
all these configurations are automatically taken into account.

\section{Motivation for the proposal from the QSC}

Our main motivation comes from the calculation of \cite{Gromov:2014eha} where the function $h(\lambda_1,\lambda_2)$ was determined for $\lambda_1=\lambda_2$ (i.e. in the ABJM theory) by comparing the localization results with the integrability-based
Quantum Spectral Curve calculation. Let us first summarize that calculation and then discuss its extension that we propose in the ABJ case.

The Quantum Spectral Curve captures the exact anomalous dimensions of all local single trace operators in ABJM model. Being based on integrability, it provides the result in terms of the effective coupling $h$. While one cannot compute a part of the spectrum via localization, the idea is to compare the localization result for the 1/6 BPS Wilson loop with the integrability prediction for the \textit{slope function} which describes the anomalous dimensions in the small spin limit. These two observables are known to be closely related in the ${\cal N}=4$ SYM theory so one can expect a similarity between them in the ABJM model as well.

The slope function $\gamma_L$ is defined as the leading coefficient in the expansion of the conformal dimension $\Delta$ of $sl(2)$ sector operators with twist $L$ at small spin $S$,
\beq
	\Delta-L-S=\gamma_L \,S+\cO(S^2)\ ,
\eeq
and it is a nontrivial function of the coupling. In \cite{Gromov:2014eha} it was computed from the QSC and the result   is written in terms of the key building blocks
\beq\la{Iab}
I_{\alpha,\beta}=\oint {dy}{} \oint  {dz} \frac{\sqrt{y-e^{4\pi h}}\sqrt{y-e^{-4\pi h}}}{\sqrt{z-e^{4\pi h}}\sqrt{z-e^{-4\pi h}}}
\frac{ y^\alpha z^\beta}{z-y}
\eeq
with the integrals going around the cut $[e^{-4\pi h},e^{4\pi h}]$.
Here the integrand has branch points at
\beq
\label{brpqsc}
	z_1=e^{4\pi h},\ z_2=e^{-4\pi h},\ z_3=\infty,\ z_4=0\;.
\eeq
The main idea of \cite{Gromov:2014eha} is to use the structural similarity between these integrals and the localization result of \cite{Marino:2009jd} for the 1/6-BPS Wilson loop.
The localization prediction can be written as
\beq
\label{Wloc}
	\langle W_{1/6}\rangle=\frac{1}{4i\pi^2  \lambda_1}I_1,\ \ I_1=\int\limits_{1/a}^a dX \arctan\sqrt\frac{(a+1/a) X-1-X^2}{(b+1/b) X+1+X^2}
\eeq
where the integrand has two branch cuts $[1/a,a]$ and $[-b,-1/b]$ formed by the condensation of eigenvalues in the matrix model of \cite{Kapustin:2009kz}
(see Fig. \ref{fig:symm} for some numerical solutions of the matrix model).
In the QSC calculation four branch points \eq{brpqsc} appear as well, suggesting they could be mapped to those of the localization result.
Indeed with a Mobius transformation we can map
\beq
	a\to e^{4\pi h},\ 1/a \to e^{-4\pi h},\ -b\to \infty,\ -1/b\to 0\ ,
\eeq
and evaluating the conformal cross-ratio of these points before and after the map gives
\beq
\label{hab3}
	h=\frac{1}{4\pi}\log\left(\frac{a b+1}{a+b}\right)\;.
\eeq

Our key observation is that exactly the same logic seem to work perfectly in the ABJ model, i.e. for non-equal couplings $\lambda_1,\lambda_2$. The localization result in this case has exactly the same form \eq{Wloc}, although it of course depends on two couplings $\lambda_1,\lambda_2$ which determine $a,b$ via the relations \eq{intl} obtained in \cite{Marino:2009jd} which we presented in section 2. Moreover, assuming the ABJ model is integrable it seems likely that the only change in the QSC would amount to using a different interpolating function $h(\lambda_1,\lambda_2)$. This is suggested by the apparent rigidity of the QSC construction together with hints from perturbative calculations and the partial weak coupling limit discussed above. In this case we would get again the same relation \eq{hab3} fixing $h(\lambda_1,\lambda_2)$.
It is important to mention that the exact form of certain effective coupling constants appearing in integrable subsectors of $\mathcal{N}=2$ SYM theories were recently conjectured, using a different approach, in \cite{Mitev:2014yba, Mitev:2015oty}.

\section{Conclusion}

In this note we proposed a conjecture for the exact all-loop interpolating function $h(\lambda_1,\lambda_2)$ in ABJ theory which should enter all integrability-based calculations in this model.
Equivalently we give a prediction for a nontrivial physical quantity - the single magnon dispersion relation.
  Our proposal is based on the same logic that allowed to fix this interpolating function for $\lambda_1=\lambda_2$ (i.e. in the ABJM model) -- namely, we map the branch points of the spectral curve arising in localization results to the branch points of the integrability-based Quantum Spectral Curve calculation. This approach turns out to work remarkably well even when the two couplings are different. Our conjecture matches all known predictions: four-loop perturbative results at weak coupling, an all-loop prediction in the limit $\lambda_1 \ll \lambda_2$, and the leading strong coupling prediction together with the expected shift of the coupling constant at strong coupling. It also has the required invariance under a Seiberg-like transformation of the couplings. The fact that all these nontrivial checks are passed is remarkable and even somewhat surprising given the compact form of our
  proposal.

One should bear in mind that it still remains an open question whether the ABJ theory is indeed integrable. We would like to emphasize the importance of a perturbative calculation which would check whether integrability persists at higher orders at weak coupling. To get significant new data one would likely need to compute the dilatation operator or anomalous dimensions at 6 loops, which though difficult may be possible to do in the superspace approach of \cite{Leoni:2010tb}.
At the same time the match of our result, which implicitly assumes integrability, with all the known data suggests that the ABJ theory is indeed integrable. The mechanism we propose allows to make several new predictions  for ABJ theory, simply by replacing $\lambda$ with $\lambda^{\text{eff}}(\lambda_1, \lambda_2)$ in the corresponding ABJM results. The replacement rule is completely determined by our proposal. We present its weak coupling expansion in Appendix \ref{appendix}.

Finally, while the approach of \cite{Gromov:2014eha} which we use in this paper is based on comparing the localization and integrability predictions for \textit{different} observables, an even more direct test would be to compute the same observable by both methods. This might be possible to do for the generalized cusp anomalous dimension if an integrability description similar to that in the ${\cal N}=4$ SYM case \cite{Drukker:2012de,Correa:2012hh,Gromov:2015dfa} is found for it.

\section*{Acknowledgements}

We thank D. Bombardelli, B. Stefanski, R. Tateo and K. Zarembo for discussions.
The research of NG and FLM leading to these results has received funding from the People Programme
(Marie Curie Actions) of the European Union's Seventh Framework Programme FP7/2007-
2013/ under REA Grant Agreement No 317089 (GATIS).
 We wish to thank
SFTC for support from Consolidated
grant number ST/J002798/1. This  project  was  partially  supported  by  the  INFN  grant  FTECP  and
UniTo-SanPaolo  Nr  TO-Call3-2012-0088 ``Modern Applications of String
Theory'' (MAST). NG and FLM would like to thank the Nordita Institute for Theoretical Physics, where a part of this work was done, for hospitality during the workshop ``Holography and Dualities 2016''.

\appendix
\section{Weak coupling map from ABJM to ABJ}\label{appendix}

Assuming our conjecture for the exact function $h(\lambda_1,\lambda_2)$ is correct, and that integrable structure of the ABJ theory differs from the one in ABJM only by changing the interpolating function, all integrability calculations done in the ABJM case would immediately be translated to the ABJ theory. In other words, the value of any observable computable by integrability in ABJ theory would be trivially obtained from the ABJM result by replacing the ABJM coupling constant $\lambda$ with an effective coupling $\lambda^\text{eff}(\lambda_1,\lambda_2)$ defined by
\beq
	h(\lambda^\text{eff},\lambda^\text{eff})=h(\lambda_1,\lambda_2)\ .
\eeq
In particular, at weak coupling we have the expansion
\beqa
	\lambda^\text{eff}&=&\sqrt{\lambda _1 \lambda _2}	
	+\sqrt{\lambda _1 \lambda _2}	(\lambda_1-\lambda_2)^2
	\[
	-\frac{\pi ^2}{12}
	+\frac{\pi ^4  \left(\lambda _1^2+32 \lambda _2 \lambda _1+\lambda _2^2\right) }{1440}
	\right. \\ \nn
	&-&
	\left.
	\frac{\pi ^6  \left(5 \lambda _1^4+178 \lambda _2 \lambda _1^3+1618 \lambda _2^2
   \lambda _1^2+178 \lambda _2^3 \lambda _1+5 \lambda _2^4\right) }{120960}
	+\dots
	\]\ \ ,
\eeqa
which is straightforwardly obtained from \eq{eq:hweak}.

\end{document}